\documentclass[conference]{IEEEtran}
\IEEEoverridecommandlockouts
\usepackage{cite}
\usepackage{xspace} 
\usepackage{amsmath,amssymb,amsfonts}
\usepackage{algpseudocode}
\usepackage{graphicx}
\usepackage{multirow}
\usepackage{textcomp}
\usepackage[table]{xcolor}
\def\BibTeX{{\rm B\kern-.05em{\sc i\kern-.025em b}\kern-.08em
    T\kern-.1667em\lower.7ex\hbox{E}\kern-.125emX}}

\newcommand{\cb}[1]{\cellcolor{blue!25} #1}
\newcommand{\crd}[1]{\cellcolor{red!25} #1}

\usepackage{hyperref}

\newif\ifarxiv
\arxivfalse

\begin{document}

\def \midd {COPERNIC}
\def \middleware {\emph{\midd}\xspace}


\title{Dynamic Service Composition Orchestrated by Cognitive Agents in Mobile \& Pervasive Computing}


\author{\IEEEauthorblockN{Oscar J. Romero}
\IEEEauthorblockA{\textit{Machine Learning Department} \\
\textit{Carnegie Mellon University}\\
5000 Forbes Avenue, Pittsburgh, USA}
}

\maketitle

\begin{abstract}
Automatic service composition in mobile and pervasive computing faces many challenges due to the complex nature of the environment. Common approaches address service composition from optimization perspectives which are not feasible in practice due to the intractability of the problem, limited computational resources of smart devices, service host's mobility, and time constraints. Our main contribution is the development of a cognitively-inspired agent-based service composition model focused on bounded rationality rather than optimality, which allows the system to compensate for limited resources by selectively filtering out continuous streams of data. The evaluation of our approach shows promising results when compared against state-of-the-art service composition models.
\end{abstract}


\vspace{-0.2cm}
\section{Introduction}
\emph{Service composition} refers to the technique of creating composite services by the aggregation of atomic services. Despite the existence of Mobile/Pervasive Computing (MPC) middleware for service composition~\cite{[5]Immonen:2014,Romero:NLS:2019,Romero:2018}, there are still some challenges that need to be tackled. Thus, we claim that service composition should:
(1) consider requests from multiple users; (2) consider resource scarcity in smart devices; (3) perform dynamic adaptation to unpredictable changes; and (4) deal with both short and long-term user's goals.
Performing service composition while taking into account the myriad of variable factors mentioned above makes the problem become intractable even for approaches that use dynamic composition and focus on optimality-based (e.g, graph-, rule-, and workflow-based) solutions, which do no consider limitations imposed by the decision-making process (specially on smart devices). Thus, we propose a cognitively-inspired, bounded-rationality-based approach so-called \middleware, which seeks satisfactory solutions rather than optimal ones, allowing composition to be performed even on resource-constrained devices.
%
\vspace{-0.5cm}
\section{Approach}
\label{sec_service_model}
\vspace{-0.3cm}

\noindent\emph{A. Preliminaries}

A \emph{concrete} service $cs_{i}$ is defined as a tuple~\cite{Balzer:2004} $\langle cs_{i}^{in}, cs_{i}^{out},$ $cs_{i}^{prec}, cs_{i}^{postc}, cs_{i}^{QoS}, cs_{i}^{ctx} \rangle$ that performs a task by acting on input data $in$ to produce output data $out$, with pre-conditions $prec$, post-conditions $postc$, Quality of Service values $QoS$, and context information $ctx$; and an \emph{abstract} service $as_i$ defined as a tuple $\langle as_{i}^{pre}, as_i^{post}, as_i^{cs} \rangle$ and realized by several concrete services that offer the same functionality ($as_i^{cs} \in \{cs_{(i,1)},...,cs_{(i,n)}\}$) such that $\forall cs_{(i,j)},$ $cs_{(i,k)} \in as_{i}^{cs} / (as_{i}^{pre} = cs_{(i,j)}^{pre} \cap cs_{(i,k)}^{pre}) \wedge (as_{i}^{post} = cs_{(i,j)}^{post} \cap cs_{(i,k)}^{post})$.

\noindent\emph{B. System Architecture}

The \middleware Agent is a cognitive module inspired by architectural principles defined by the Common Model of Cognition (CMC)\cite{llc:2017,Romero:ADL:2018}, a computational model that captures a consensus about the structures and processes that are similar to those found in human cognition. Cognitive modules are (see detailed description in~\cite{Romero:2019}):
\emph{\textbf{1. Perception:}} it perceives agent's current state by processing both external (e.g., user requests) and internal (i.e., user's context, sensor readings, and QoS) sensory inputs. It outputs a set of percepts (symbolic representation units) such $P = \{p_{1}...p_{n}\}$ $|$ $\forall p_{i} \in PR$, where $PR$ is a set of premises such that $as^{pre} \cup cs^{pre} \subseteq PR$.
\emph{\textbf{2. Working Memory (WM):}} WM holds previous percepts not yet decayed away, and local associations from declarative memories that are combined with the percepts to understand the current state of the service composition. WM defines a limited storage capacity and a recency-based decay function that keeps active a limited number of units, expressed as a base-level activation function. Contents of WM are used as inputs for service matching, i.e., $w_{i} \subseteq (as^{pre} \cup cs^{pre}) \subseteq PR$.
\emph{\textbf{3. Declarative Memories:}} the WM cues the declarative memories (i.e., \emph{Episodic Memory (EM)} that retrieves information about services' historic performance, context, etc., and \emph{Semantic Memory (SM)} that retrieves service descriptions, user preferences, etc.) and stores local associations. EM is a content-addressable associative memory represented through a Sparse Distributed Memory; and SM is implemented using a Slipnet, an activation passing semantic network. This module outputs a set of premises $D = \{d_{1}...d_{n}\}$ $|$ $\forall d_{i} \in PR$.
\emph{\textbf{4. Selective Attention (SA):}} SA filters out a continuous stream of content from WM so the agent only focuses on the most relevant information needed for matching abstract services. Goals are decomposed and abstract services compete and cooperate among them in order to get the focus of attention.
SA uses a Behavior Network (BN)~\cite{maes,Romero:2011}, a hybrid system that integrates both spreading activation dynamics and a symbolic, structured representation. Each behavior of the BN maps to a single abstract service $as$, and \emph{``service discovery''} emerges from the  activation$/$inhibition dynamics among all services. Abstract services (behaviors) distinguish between expected/non-expected (positive/negative) postconditions (in terms of an ``add'' and a ``delete'' list) and define a level of activation $as_{i}^{\alpha}$.
%
%
Also, the model defines 5 parameters to tune the global behavior of the network: $\pi$ is the mean level of activation, $\theta$ is the threshold for becoming active, $\phi$ is the amount of activation energy a WM unit injects into the network, $\gamma$ is the amount of energy a goal injects into the network, and $\delta$ is the amount of activation energy a protected goal takes away from the network. 
\emph{\textbf{5. Procedural Memory (PM):}} PM defines a set of heuristics to: 1) discover concrete services based on QoS attributes; and 2) adjust the BN parameters to make the global behavior be more adaptive. PM applies the following heuristics~\cite{Romero:2011} to keep the balance between: (1) goal-orientedness vs. situation-orientedness, $\gamma > \phi$; (2) deliberation vs. reactivity, $\phi > \gamma \wedge \phi > \theta$; and (3) bias towards ongoing plan vs. adaptivity, $\phi > \pi > \gamma$. The values are dynamically adapted using a utility-based learning mechanism. 
\noindent\emph{\textbf{6. Action Selection (AS):}} AS processes both internal actions (such as goal setting) and and external actions (such as triggering a device's effector/actuator, and invoking the discovery mechanism to execute concrete services).
\emph{\textbf{7. Cognitive Cycle:}} unlike traditional approaches that create upfront composition plans which are prone to inadaptability, in our approach, plans emerge from the interaction of cascading sequences of \textbf{cognitive cycles} corresponding to perception-action loops (modules 1-6) where compositional conditions are validated and reasoning and planning take place. This contribution allows service composition to be more reactive, robust, and adaptive to dynamic changes while composition plans are generated on-the-fly by using minimal resources as a result of filtering out a continuous stream of information.


\vspace{-0.35cm}
\section{Evaluation}
\label{sec_evaluation}
\vspace{-0.2cm}

We used the NS-3 simulator to compare the performance of \middleware\footnote{\textcolor{blue}{\url{https://github.com/ojrlopez27/copernic}}} against two state-of-the-art decentralized service composition models: GoCoMo, a goal-driven service model based on a decentralized heuristic backward-chaining planning algorithm~\cite{Chen2018}; and CoopC, a decentralized goal-driven cooperative composition model that does not support runtime composite service adaptation~\cite{Furno:2013}. We modified service density (sparse (SD-S): 20, medium (SD-M): 40, dense (SD-D): 60); composition length (5 services (CL-5) or 10 services (CL-10)); and node mobility (slow (M-S): 0-2m/s, medium (M-M): 2-8m/s, and fast (M-F): 8-13m/s); and we measured 3 different metrics: composition time (\textbf{CT} in seconds), average memory used during the composition (\textbf{MU} in Kb), and a planning failure rate \textbf{PFR} (\# of failed planning processes / \# of all the issued requests).
%
%
Table~\ref{tab:flexibility} shows the results (blue/red cells are the best/worst measurements for each category, respectively). 
In particular, GoCoMo's failure rate was lower than \middleware (12-38\%) when the mobility was slow. This difference dropped to 7-13\% in fast-mobility high-density scenarios because \middleware is less sensitive to mobility changes thanks to service information is stored in the WM and gradually fades away, which means that it can still be accessible even when a service disappears and reappears later in time, 
%
%
allowing the service to promptly participate again in the composition without producing significant planning failures. In comparison with CoopC, \middleware got 12-25\% less failures due to CoopC does not support runtime adaptation and poorly handles mobility changes. Regarding composition time, \middleware tailored composite services up to 42\% and 71\% faster than GoCoMo and CoopC, respectively; and it used up to 72\% and 84\% less memory than GoCoMo and CoopC, respectively. The reason for this significant reduction is that \middleware is continuously filtering out the stream of incoming information, which keeps it into reasonable margins of resources usage, despite of the dynamism of the environment. 
It is worth noting that \middleware did not show a significant difference in memory usage when using a composition length of either 5 or 10 services (-4\% - 11\%) in comparison with GoCoMo  (60\% - 190\%) and CoopC (157\% - 201\%), which suggests that our approach could be smoothly scaled up.

\begin{table}[htb]
\centering
\vspace{-0.4cm}
\caption{Flexibility of Service Composition}
\vspace{-0.2cm}
\label{tab:flexibility}
\resizebox{\columnwidth}{!}{%
\begin{tabular}{l|l|l|r|r|r|r|r|r|r|r|r} 
\hline
\multicolumn{3}{l|}{}                                                                                                                                                                & \multicolumn{3}{c|}{\textbf{M-S} }                                                                                                                                        & \multicolumn{3}{c|}{\begin{tabular}[c]{@{}c@{}}\textbf{M-M}\\ \end{tabular}}                                                                                              & \multicolumn{3}{c}{\textbf{M-F} }                                                                                                                                          \\ 
\cline{4-12}
\multicolumn{1}{l}{}                                                        & \multicolumn{1}{l}{} &                                                                                 & \multicolumn{1}{c|}{\textbf{SD-S} }                     & \multicolumn{1}{c|}{\textbf{SD-M} }                    & \textbf{SD-D}                                          & \multicolumn{1}{c|}{\textbf{SD-S} }                     & \multicolumn{1}{c|}{\textbf{SD-M} }                    & \textbf{SD-D}                                          & \multicolumn{1}{c|}{\textbf{SD-S} }                     & \multicolumn{1}{c|}{\textbf{SD-M} }                    & \textbf{SD-D}                                           \\ 
\hline
\multirow{2}{*}{\textbf{COPERNIC} }                                         & \textbf{CL-5}        & \begin{tabular}[c]{@{}l@{}}\textbf{PFR (\%)}\\\textbf{CT (sec)}\\\textbf{MU (Kb)} \end{tabular} & \begin{tabular}[c]{@{}l@{}}18.2\\0.9\\\cb{63} \end{tabular}  & \begin{tabular}[c]{@{}l@{}}3.7\\\cb{0.5}\\81 \end{tabular}  & \begin{tabular}[c]{@{}l@{}}1.1\\0.8\\93 \end{tabular}  & \begin{tabular}[c]{@{}l@{}}17.5\\1.1\\67 \end{tabular}  & \begin{tabular}[c]{@{}l@{}}1.4\\1.2\\86 \end{tabular}  & \begin{tabular}[c]{@{}l@{}}1.4\\1.2\\93 \end{tabular}  & \begin{tabular}[c]{@{}l@{}}21.1\\1.1\\73 \end{tabular}  & \begin{tabular}[c]{@{}l@{}}3.3\\1.4\\88 \end{tabular}  & \begin{tabular}[c]{@{}l@{}}1.1\\1.4\\98 \end{tabular}   \\ 
\cline{2-12}
                                                                            & \textbf{CL-10}       & \begin{tabular}[c]{@{}l@{}}\textbf{PFR (\%)}\\\textbf{CT (sec)}\\\textbf{MU (Kb)} \end{tabular} & \begin{tabular}[c]{@{}l@{}}17.8\\1.2\\70 \end{tabular}  & \begin{tabular}[c]{@{}l@{}}3.7\\0.6\\86 \end{tabular}  & \begin{tabular}[c]{@{}l@{}}1.1\\0.8\\92 \end{tabular}  & \begin{tabular}[c]{@{}l@{}}17.7\\1.2\\70 \end{tabular}  & \begin{tabular}[c]{@{}l@{}}1.5\\1.2\\73 \end{tabular}  & \begin{tabular}[c]{@{}l@{}}0.5\\1.1\\85 \end{tabular}  & \begin{tabular}[c]{@{}l@{}}19.7\\1.2\\78 \end{tabular}  & \begin{tabular}[c]{@{}l@{}}3.8\\1.3\\89 \end{tabular}  & \begin{tabular}[c]{@{}l@{}}1.4\\1.9\\94 \end{tabular}   \\ 
\hline
\multirow{2}{*}{\begin{tabular}[c]{@{}c@{}}\textbf{GoCoMo}\\ \end{tabular}} & \textbf{CL-5}        & \begin{tabular}[c]{@{}l@{}}\textbf{PFR (\%)}\\\textbf{CT (sec)}\\\textbf{MU (Kb)} \end{tabular} & \begin{tabular}[c]{@{}l@{}}13.1\\1.3\\79 \end{tabular}  & \begin{tabular}[c]{@{}l@{}}3.3\\0.7\\93 \end{tabular}  & \begin{tabular}[c]{@{}l@{}}0.6\\0.9\\112 \end{tabular} & \begin{tabular}[c]{@{}l@{}}16.1\\1.3\\78 \end{tabular}  & \begin{tabular}[c]{@{}l@{}}1.2\\1.4\\93 \end{tabular}  & \begin{tabular}[c]{@{}l@{}}\cb{0.3}\\1.4\\110 \end{tabular} & \begin{tabular}[c]{@{}l@{}}18.0\\1.3\\80 \end{tabular}  & \begin{tabular}[c]{@{}l@{}}3.1\\1.3\\94 \end{tabular}  & \begin{tabular}[c]{@{}l@{}}0.9\\1.4\\114 \end{tabular}  \\ 
\cline{2-12}
                                                                            & \textbf{CL-10}       & \begin{tabular}[c]{@{}l@{}}\textbf{PFR (\%)}\\\textbf{CT (sec)}\\\textbf{MU (Kb)} \end{tabular} & \begin{tabular}[c]{@{}l@{}}16.2\\2.1\\213 \end{tabular} & \begin{tabular}[c]{@{}l@{}}2.3\\2.2\\273 \end{tabular} & \begin{tabular}[c]{@{}l@{}}0.8\\2.2\\314 \end{tabular} & \begin{tabular}[c]{@{}l@{}}24.7\\2.2\\201 \end{tabular} & \begin{tabular}[c]{@{}l@{}}1.1\\2.3\\287 \end{tabular} & \begin{tabular}[c]{@{}l@{}}0.4\\2.3\\308 \end{tabular} & \begin{tabular}[c]{@{}l@{}}22.1\\2.3\\221 \end{tabular} & \begin{tabular}[c]{@{}l@{}}3.5\\2.4\\286 \end{tabular} & \begin{tabular}[c]{@{}l@{}}1.3\\2.4\\345 \end{tabular}  \\ 
\hline
\multirow{2}{*}{\begin{tabular}[c]{@{}c@{}}\textbf{CoopC}\\ \end{tabular}}  & \textbf{CL-5}        & \begin{tabular}[c]{@{}l@{}}\textbf{PFR (\%)}\\\textbf{CT (sec)}\\\textbf{MU (Kb)} \end{tabular} & \begin{tabular}[c]{@{}l@{}}16.2\\1.8\\114\end{tabular}  & \begin{tabular}[c]{@{}l@{}}2.4\\1.9\\245\end{tabular}  & \begin{tabular}[c]{@{}l@{}}0.8\\1.9\\367\end{tabular}  & \begin{tabular}[c]{@{}l@{}}21.9\\1.9\\121\end{tabular}  & \begin{tabular}[c]{@{}l@{}}1.3\\1.8\\239\end{tabular}  & \begin{tabular}[c]{@{}l@{}}2.3\\2.1\\353\end{tabular}  & \begin{tabular}[c]{@{}l@{}}24.5\\1.9\\117\end{tabular}  & \begin{tabular}[c]{@{}l@{}}3.7\\2.1\\275\end{tabular}  & \begin{tabular}[c]{@{}l@{}}1.2\\2.2\\359\end{tabular}   \\ 
\cline{2-12}
                                                                            & \textbf{CL-10}       & \begin{tabular}[c]{@{}l@{}}\textbf{PFR (\%)}\\\textbf{CT (sec)}\\\textbf{MU (Kb)} \end{tabular} & \begin{tabular}[c]{@{}l@{}}24.0\\4.1\\325\end{tabular}  & \begin{tabular}[c]{@{}l@{}}2.3\\4.2\\476\end{tabular}  & \begin{tabular}[c]{@{}l@{}}1.3\\4.2\\593\end{tabular}  & \begin{tabular}[c]{@{}l@{}}25.2\\4.5\\332\end{tabular}  & \begin{tabular}[c]{@{}l@{}}2.4\\4.7\\488\end{tabular}  & \begin{tabular}[c]{@{}l@{}}1.2\\4.9\\605\end{tabular}  & \begin{tabular}[c]{@{}l@{}}\crd{31.8}\\5.0\\345\end{tabular}  & \begin{tabular}[c]{@{}l@{}}4.2\\5.1\\497\end{tabular}  & \begin{tabular}[c]{@{}l@{}}1.6\\\crd{5.5}\\\crd{657}\end{tabular}   \\
\hline
\end{tabular}
}
\end{table}

\vspace{-0.5cm}
\section{Conclusions and Future Work}
\label{sec_conclusions}
\vspace{-0.2cm}

We propose a cognitive model that efficiently and dynamically orchestrates distributed services under highly changing conditions. Our approach focuses on bounded rationality rather than optimality, allowing the system to compensate for limited resources by filtering out a continuous stream of incoming information. We tested our model against state-of-the-art service composition models while modifying mobility, service density and composition complexity features, and the results were promising demonstrating that our approach may be suitable for MPC environments where resources are scarce. 
\vspace{-0.2cm}

\bibliographystyle{IEEEtranS}
\bibliography{main} 

\begin{thebibliography}{10}
\providecommand{\url}[1]{#1}
\csname url@samestyle\endcsname
\providecommand{\newblock}{\relax}
\providecommand{\bibinfo}[2]{#2}
\providecommand{\BIBentrySTDinterwordspacing}{\spaceskip=0pt\relax}
\providecommand{\BIBentryALTinterwordstretchfactor}{4}
\providecommand{\BIBentryALTinterwordspacing}{\spaceskip=\fontdimen2\font plus
\BIBentryALTinterwordstretchfactor\fontdimen3\font minus
  \fontdimen4\font\relax}
\providecommand{\BIBforeignlanguage}[2]{{%
\expandafter\ifx\csname l@#1\endcsname\relax
\typeout{** WARNING: IEEEtranS.bst: No hyphenation pattern has been}%
\typeout{** loaded for the language `#1'. Using the pattern for}%
\typeout{** the default language instead.}%
\else
\language=\csname l@#1\endcsname
\fi
#2}}
\providecommand{\BIBdecl}{\relax}
\BIBdecl

\bibitem{Balzer:2004}
S.~Balzer, ``Bridging the gap between abstract and concrete services a semantic
  approach for grounding owl-s,'' in \emph{Semantic WS}, 2004.

\bibitem{Chen2018}
N.~Chen and S.~Clarke, ``Goal-driven service composition in mobile and
  pervasive computing,'' \emph{Services Computing}, vol.~11, no.~1, 2018.

\bibitem{Furno:2013}
A.~Furno, ``Efficient cooperative discovery of service compositions in
  unstructured p2p networks,'' in \emph{Parallel Processing}, 2013.

\bibitem{[5]Immonen:2014}
A.~Immonen and D.~Pakkala, ``A survey of methods and approaches for reliable
  dynamic service compositions,'' \emph{SOA}, vol.~8, 2014.

\bibitem{llc:2017}
J.~Laird, ``A {Standard} {Model} of the {Mind}: Toward a common computational
  framework across artificial intelligence, cognitive science, neuroscience,
  and robotics,'' \emph{AI}, no.~4, 2017.

\bibitem{maes}
P.~Maes, ``How to do the right thing,'' \emph{Connection Science}, 1989.

\bibitem{Romero:2011}
O.~J. Romero, ``An evolutionary behavioral model for decision making,''
  \emph{Adaptive Behavior}, vol.~19, no.~6, pp. 451--475, 2011.

\bibitem{Romero:ADL:2018}
------, ``{{CogArch-ADL}: Toward a Formal Description of a Reference
  Architecture for the {Common Model of Cognition}},'' \emph{PCS}, 2018.

\bibitem{Romero:2019}
------, ``{Dynamic Service Composition Orchestrated by Cognitive Agents in
  Mobile \& Pervasive Computing},'' in \emph{AIMS}, 2019, p. In press.

\bibitem{Romero:2018}
O.~J. Romero and S.~Akoju, ``An efficient mobile-based middleware architecture
  for building robust, high-performance apps,'' in \emph{ICSA}, 2018.

\bibitem{Romero:NLS:2019}
O.~J. Romero and A.~Dangi, ``{NLSC: Unrestricted Natural Language-based Service
  Composition through Sentence Embeddings},'' in \emph{SCC}, 19.

\end{thebibliography}

\end{document}